\documentclass[a4paper,twoside,notoc,11pt]{JHEP3}

\usepackage{epsfig,multicol}
\usepackage{delarray,amsmath,bbm}

\newcommand\fverb{\setbox\pippobox=\hbox\bgroup\verb}
\newcommand\fverbdo{\egroup\medskip\noindent%
                              \fbox{\unhbox\pippobox}\ }
\newcommand\fverbit{\egroup\item[\fbox{\unhbox\pippobox}]}
\newbox\pippobox
\newcommand{\nn}{\nonumber}
\newcommand{\beq} {\begin{equation}}
\newcommand{\eeq} {\end{equation}}
\newcommand{\beqa} {\begin{eqnarray}}
\newcommand{\eeqa} {\end{eqnarray}}
\newcommand{\mrm}[1] {{\mathrm{#1}}}

\newcommand{\ie}{{\it i.e.}}
\newcommand{\eg}{{\it e.g.}}

\newcommand{\wrt}{{\it wrt.\ }}

\newcommand{\gev}{{\mrm{GeV}}}

\newcommand{\lqcd}{\Lambda_{QCD}}

\newcommand{\ieps}{i\varepsilon}
\newcommand{\order}[1]{${\cal O}\left(#1 \right)$}

\newcommand{\eq}[1]{(\ref{#1})}

\newcommand{\inv}[1]{\frac{1}{#1}}

\newcommand{\bs}[1]{\boldsymbol{#1}}

\newcommand{\im}{{\rm Im}}

\newcommand{\mM}{\mathcal{M}}
\newcommand{\mN}{\mathcal{N}}

\newcommand{\kvec}{{\bs{k}}_\perp}
\newcommand{\lvec}{{\bs{\ell}}_\perp}

\newcommand{\lvect}{\bs{\ell}_{2\perp}}
\newcommand{\notvec}{\bs{0}_\perp}
\newcommand{\kt}{k_\perp}

\newcommand{\ellt}{\ell_\perp}

\newcommand{\lto}{\ell_{1 \perp}}
\newcommand{\ltt}{\ell_{2 \perp}}

\newcommand{\kex}[1]{k_\perp e^{{#1}i\phi}}
\newcommand{\qex}[1]{\ell_\perp e^{{#1}i\psi}}

\newcommand{\lext}[1]{\ltt e^{{#1}i\tau_2}}

\newcommand{\mcal}{\mathcal{M}}

\newcommand{\halft}{{\textstyle \frac{1}{2}}}

\newcommand{\gsim}{\gtrsim}

\title{\center{Single Spin Asymmetry at Large $x_F$ and $\bs{k_\perp}$}}
\author{Paul Hoyer and Matti J\"arvinen\\
              Department of Physical Sciences and Helsinki Institute of
              Physics\\
              \ POB 64, FIN-00014 University of Helsinki, Finland \\
              E-mails: \email{paul.hoyer@helsinki.fi, Matti.O.Jarvinen@Helsinki.fi}}
\preprint{HIP-2006-51/TH \\  
\hepph{0611293} \\ \today
}
\abstract{The large single spin asymmetries observed at high momentum fractions $x_F$ and transverse momenta $\kt$ in $p^\uparrow p \to \pi(x_F,\kvec)+X$ as well as in $pp \to \Lambda^\uparrow(x_F,\kvec)+ X$ suggest that soft helicity flip processes are coherent with hard scattering. Such coherence can be maintained if $x_F \to 1$ as $\kt \to \infty$, while $\kt^2(1-x_F) \sim \lqcd^2$ stays fixed. The entire hadron wave function, rather than a single quark, then contributes to the scattering process. Analogous coherence effects have been seen experimentally in the Drell-Yan process at high $x_F$. We find that the $p^\uparrow p \to \pi(x_F,\kvec)+X$ production amplitudes have large dynamic phases and that helicity flip contributions are unsuppressed in this limit, giving rise to potentially large single spin asymmetries.
}


\begin{document}

\noindent{\bf 1. Coherence effects at high $x_F$}\vspace{.5cm}

In this paper we suggest a possible dynamical explanation of the large single spin asymmetries (SSA) observed in polarized hadron scattering, $p^\uparrow p \to \pi(x_F,\kvec)+X$ \cite{Adams:1991rw,Adams:2003fx} and $pp \to \Lambda^\uparrow(x_F,\kvec)+ X$ \cite{Bunce:1976yb,Lundberg:1989hw}. At the highest measured longitudinal momentum fractions $x_F \simeq 0.8$ the $p^\uparrow p \to \pi X$ asymmetry rises to $A_N \sim 0.4$, and increases for transverse momenta above $\kt = 0.7$ GeV. The $\Lambda$ polarization reaches $P_\Lambda \gsim 0.3$ at $x_F \simeq 0.8$, and does not decrease with transverse momentum up to the highest measured value $\kt \simeq 3.5$ GeV. These asymmetries are an order of magnitude larger than those observed in DIS ($ep^\uparrow \to e+ X$) \cite{Airapetian:2004tw}. This motivates us to consider a dynamics where the entire polarized hadron wave function contributes coherently to the production process.

In the standard leading twist (LT) limit of $p p \to \pi(x_F,\kt)+ X$,
\beq\label{ltlimit}
{\rm LT:}\ \ \ \kt \to \infty \ \ \ {\rm at\ fixed}\ \ x_F < 1
\eeq
each projectile parton (mostly quarks at high $x_F$) contributes incoherently to the process. In order to produce a pion with $x_F=0.8$ there must be a quark which carries a momentum fraction $x \gsim 0.9$ of the proton, and which after its hard scattering transfers a fraction $z \gsim 0.9$ of its momentum to the pion. These stringent requirements imply a very small production cross section. In fact, QCD at leading twist was found to underestimate the pion production cross section measured by E704 \cite{Adams:1991rw} by an order of magnitude at high $x_F$ \cite{Bourrely:2003bw}. On the other hand, the cross section measured at lower $x_F$ and higher $\kt$ by STAR \cite{Adams:2003fx} is consistent with LT QCD. Our present discussion focusses on the E704 kinematics. 

Proton wave function components where one quark carries most of the momentum have a short life-time even though they do not involve large transverse momenta. The energy difference $\Delta E$ between a proton of large momentum $p$ and a Fock state with partons carrying longitudinal momentum fractions $x_i$ and transverse momenta $k_{i\perp}$ is
\beq\label{deltae}
2p\Delta E = m_p^2 - \sum_i \frac{k_{i\perp}^2+m_i^2}{x_i}
\eeq
where $\sum_i x_i = 1$. The energy difference is large, \ie, the lifetime of the Fock state is short, if any $x_i \simeq 1$ since then $x_j \simeq 0\ (j\neq i)$. For the leading twist approximation to be valid when such Fock states scatter the hard scale $\kt$ must be much larger than the intrinsic scale \eq{deltae}, \ie, $\kt^2 \gg \lqcd^2/(1-x_i)$.

As shown by Berger and Brodsky (BB) \cite{Berger:1979du}, the angular distribution of the muon pair in the Drell-Yan process $\pi N \to \mu\mu(x_F)+ X$ provides a measure of the coherence effects which set in at high $x_F$. When the intrinsic hardness of the contributing pion Fock states becomes comparable to the virtuality $Q^2$ of the photon the angular distribution of the muons, which is $1+\cos^2\theta$ at leading twist, turns into $\sin^2\theta$. In effect, the virtual photon couples coherently to the pion Fock states which contribute at high $x_F$. Thus the helicity of the pion (which is zero) is transferred to the virtual photon, giving a $\sin^2\theta$ decay distribution. This phenomenon was subsequently observed in the Drell-Yan data \cite{Anderson:1979xx,Falciano:1986wk}. The change of angular distribution occurs at $x_F \simeq 0.7$ for $Q^2 \simeq 20\ \gev^2$ in E615 \cite{Anderson:1979xx}. 

We shall explore the relevance of the BB dynamics for the E704 SSA effect. A large asymmetry is more understandable if the pion is created coherently from the proton wave function, which contains the entire spin information. The E615 observation lead us to expect considerable coherence effects at $x_F \simeq 0.8$ of the pion in E704, given its moderate $\kt\simeq 1$ GeV. The leading twist contribution to the E704 data is suppressed also by the quark fragmentation ($z \gsim 0.9$), whereas there is no fragmentation in the Drell-Yan process.

Thus we consider $p^\uparrow p \to \pi(x_F,\kvec)+X$ in the limit\footnote{In the BB Drell-Yan analysis \cite{Berger:1979du} the coherence sets in already at fixed $Q^2(1-x_F)^2$. This is because of a helicity mismatch in the leading twist parton distribution $f_{q/\pi}$, whereas helicity is conserved in $\pi \to \gamma_L$. The same scaling behavior was noted for $ep\to e+X$ and $e^+e^-\to h+X$ in \cite{Farrar:1975yb}.}
\beq\label{bblimit}
{\rm BB:}\ \ \ \kt \to \infty \ \ \ {\rm at\ fixed}\ \ \kt^2(1-x_F) \sim \lqcd^2
\eeq
The contribution of a quark with large transverse momentum $\kt$ and $x \simeq x_F \simeq 1$ to the energy difference \eq{deltae} is then of the same order as the contribution of quarks carrying transverse momenta of \order{\lqcd} and momentum fractions of \order{1-x_F}. Thus the hard and soft parts of the wave function remain coherent in the \order{1/\kt} (proper) time scale of the hard process. Soft interactions of the low momentum quarks then influence the hard process, as was demonstrated for quarkonium production at high $x_F$ \cite{Brodsky:1991dj}.

The standard theoretical framework based on the QCD twist expansion is inapplicable in the BB limit \eq{bblimit}. We construct an explicit perturbative example which demonstrates the possibility of a spin asymmetry $A_N$, analogous to the demonstration of an SSA at leading twist in DIS by Brodsky, Hwang and Schmidt (BHS) \cite{Brodsky:2002cx}. We pay particular attention to the almost paradoxical requirement, pointed out long ago \cite{Kane:1978nd}, that a spin asymmetry of \order{1} in a hard process requires the helicity flip and non-flip amplitudes to be of similar size and to have a large dynamical phase difference. This is seen by expressing the transverse spin amplitudes $\mcal_{\updownarrow,\{\sigma\}}$ 
(with spin in the $y$ direction) in terms of helicity amplitudes $\mcal_{\leftrightarrow,\{\sigma\}}$. In the case of $p^\uparrow p \to \pi(x_F,\kvec)+X$ we define
\beq \label{asymm}
 A_N(x_F,\kt) \cos\phi \equiv \frac{\sum_{\{\sigma\}}\left[|\mcal_{\uparrow,\{\sigma\}}|^2-|\mcal_{\downarrow,\{\sigma\}}|^2 \right]}{\sum_{\{\sigma\}}\left[|\mcal_{\uparrow,\{\sigma\}}|^2+|\mcal_{\downarrow,\{\sigma\}}|^2  \right]} 
= \frac{2 \sum_{\{\sigma\}} \im\left[\mcal_{\leftarrow,\{\sigma\}}^*\mcal_{\rightarrow,\{\sigma\}}\right]}{\sum_{\{\sigma\}}\left[|\mcal_{\rightarrow,\{\sigma\}}|^2+|\mcal_{\leftarrow,\{\sigma\}}|^2  \right]}
\eeq 
where $\phi$ is the azimuthal angle of the outgoing pion. Thus $A_N(x_F,\kt)$ refers to the maximal asymmetry at $\phi=0,\pi$ when the pion transverse momentum is perpendicular to the spin direction. The helicities $\{\sigma\}$ of all particles except the polarized one are the same in all amplitudes. In addition to summing over $\{\sigma\}$, the numerator and the denominator of \eq{asymm} need to be separately integrated over the phase space of the unobserved particles of the final state.

\EPSFIGURE[h]{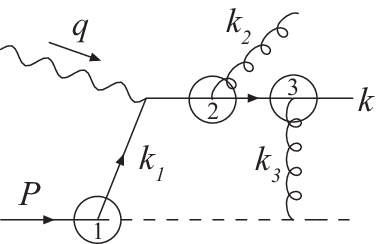,width=0.4\columnwidth}{Diagram illustrating the generation of an SSA in semi-inclusive DIS. At leading order in $1/q^2$ a large $\kt$ and/or helicity flip can be generated at any of the three vertices marked by a circle. The dashed line is a scalar particle.
\label{fig1}}

\vspace{1cm}

\noindent{\bf 2. Sources of a single spin asymmetry at high $\kt$ and fixed $x_F$}
\vspace{.5cm}

Let us consider semi-inclusive DIS ($ep\to eq+X$) to illustrate the suppression of $A_N$ \cite{Kane:1978nd} in quark (or hadron) production at large transverse momentum $\kt$. To generate the dynamical phase required by \eq{asymm} we need a loop diagram. In the BHS diagram\footnote{In the original BHS work $\kt \sim \lqcd$ was small (the hard scale in DIS is given by $Q^2$), and there was no gluon radiation $k_2$.} \cite{Brodsky:2002cx} of Fig.~1 transverse momentum and helicity flip can be generated at any of the three vertices marked by a circle.

Amplitudes where the helicity flip and large transverse momentum are generated at the same vertex are proportional to the small current quark mass and thus cannot give rise to a sizeable $A_N$. 

Helicity flip at a soft vertex (say, vertex 1 in Fig.~1) is proportional to $\lqcd/k_{1\perp}$. With $k_{1\perp}\sim \lqcd$ the flip and non-flip amplitudes are of the same order (as in the original BHS model). Then the azimuthal distribution of a hard gluon $k_2$ radiated at vertex 2 is, however, independent of the quark helicity to leading order in $k_{1\perp}/k_{2\perp}$. The azimuthal distribution of the soft momentum $\bs{k}_1$ leaves a residual effect $A_N \sim \lqcd/k_{2\perp}$ due to ``trigger bias'', as first pointed out by Sivers \cite{Sivers:1989cc}. It has in fact proved possible to fit the E704 data using the Sivers effect \cite{Anselmino:1994tv}. A recent study in the (presumably equivalent \cite{Ji:2006ub}) 
twist-3 operator approach \cite{Efremov:1981sh,Qiu:1991pp} found it, on the other hand, difficult to account for the E704 data \cite{Kouvaris:2006zy}. The underestimate  of the E704 cross section \cite{Bourrely:2003bw} and the increase of $A_N$ above $\kt = 0.7$ GeV casts doubt on the Sivers effect as the source of the large $A_N$ seen at high $x_F$ .

In the previous example the soft emission at vertex 1 was incoherent with, and thus decoupled from, the hard dynamics at vertex 2 since $k_{1\perp} \ll k_{2\perp}$. On the other hand, when the soft process is coherent with the hard one it directly affects the angular distribution of the hard emission. Due to time dilation the soft rescattering of the struck quark at vertex 3 is coherent with the hard emission, which makes it relevant to the phase difference of the amplitudes in \eq{asymm}. This was the essential observation of BHS \cite{Brodsky:2002cx}, allowing a non-vanishing $A_N$ at leading order in $Q^2$. However, quark helicity flip in the high energy $(\propto \nu)$ rescattering at vertex 3 is $\propto m_q/\nu$ for the Born contribution of Fig.~1. Helicity flip occurs to leading order in $\nu$ only through the anomalous magnetic moment of the quark, which perturbatively arises from (soft) gluon loop vertex corrections. The long formation time of a sizeable anomalous magnetic moment makes spin flip incoherent with the hard process. We shall not pursue here the possibility \cite{Hoyer:2005ev} that this argument might fail due to QCD vacuum effects.

\vspace{1cm}

\noindent{\bf 3. Single spin asymmetry at $\kt^2(1-x_F) \sim \lqcd^2$}
\vspace{.5cm}

As we discussed in section 1, there are reasons to believe that the BB limit \eq{bblimit} is relevant for the E704 kinematics. We now demonstrate that a sizeable $A_N$ naturally arises in this limit. Due to the scaling $\kt^2 \sim \lqcd^2/(1-x_F)$ the spin asymmetry extends to higher transverse momenta than in the fixed $x_F$ limit discussed in section 2, where $A_N \sim \lqcd/\kt$ is expected.

In $p^\uparrow p \to \pi(x_F,\kvec)+X$ the pion shares one valence quark with the proton. For $x_F \to 1$ the other proton constituents must be stopped; they may carry momentum fractions only of \order{1-x_F}. The energy difference \eq{deltae} is minimal when the stopped partons carry transverse momenta of \order{\lqcd}, in which case they cannot balance the large transverse momentum $\kvec$ of the pion. Thus we expect that the fast valence quark gains its transverse momentum from a hard collision with a target parton. 

\EPSFIGURE[t]{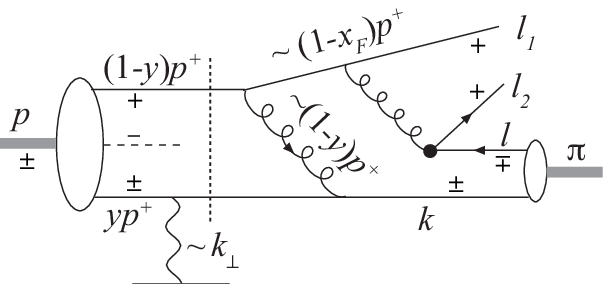,width=0.8\columnwidth}{Feynman diagram contributing to $pp \to \pi+X$ in the limit \eq{bblimit}. One of the proton valence quarks (dashed line) is assumed to have already transferred its momentum to the other two. The valence quark $yp^+$ gains a large transverse momentum $\kt$ from scattering on a target parton, as well as most of the momentum $(1-y)p^+$ of the remaining valence quark. The quarks $\ell_1,\ell_2$ and antiquark $\ell$ have small momentum fractions of \order{(1-x_F)} and transverse momenta of \order{\lqcd}. Helicities are indicated by $\pm$ signs. In one of the helicity amplitudes a helicity flip occurs at the vertex indicated by the black dot.
\label{fig2}}

We consider the scenario shown in the Feynman diagram of Fig.~2. The projectile proton enters with a large momentum along the $z$-axis, $p=(p^+,0,\bs{0}_\perp)$, where $p^\pm = p^0 \pm p^z$. In order to simplify the diagram we assume that one of the proton quarks (dashed line) has already transferred its longitudinal momentum to the other two, which thus carry momentum fractions $y$ and $1-y$ of \order{\halft}. The quark with momentum $yp^+$ scatters on the target, acquring a transverse momentum of \order{\kvec}. The other quark transfers nearly all of its longitudinal momentum $\sim (1-y)p^+$ to the one carrying $yp^+$. At this point we have created a short-lived state, with a fast ($\sim p^+$) quark of high transverse momentum ($\kvec$) and two slow ($\sim (1-x_F)p^+$) quarks of low transverse momentum ($\sim \lqcd$). All three quarks contribute equally to the energy difference \eq{deltae} and thus remain fully coherent. The proton wave function is integrated over its intrinsic transverse momentum up to the hard scale $\kt$, giving its distribution amplitude \cite{Lepage:1980fj,Mueller:1981sg,Brodsky:24wx}.

Due to their reduced time dilation, the soft interactions of the slow quarks occur at the same time-scale as the hard scattering of the fast quark,
\beq\label{tscales}
t_{soft} \simeq \inv{\lqcd} \frac{(1-x_F)p^+}{\lqcd} \sim \frac{p^+}{\kt^2} \sim \inv{\kt}\frac{p^+}{\kt} \simeq t_{hard}
\eeq
In particular, a slow quark may, within the overall coherence time, create the quark-antiquark pair shown in Fig.~2, with momenta $\ell_2^+,\ell^+ \sim (1-x_F)p^+$ and transverse momenta of \order{\lqcd}. Finally, the pion is formed at the common time-scale \eq{tscales} through a gluon exchange (not shown) from the fast quark ($k$) to the antiquark ($\ell$), which gives them an equal share the pion momentum. Similarly to the proton, the pion wave function enters via its valence distribution amplitude.

Several Feynman diagrams contribute to the pion production process. The one shown in Fig.~2 should be representative, \ie, no essential cancellations with the other diagrams are expected. This is the leading contribution to pion production from unpolarized protons in the BB limit \eq{bblimit}, and it should account for the difference between the leading twist cross section and the one measured by E704. We shall evaluate the diagram of Fig.~2 below, but since it contains soft subprocesses the perturbative result is not quantitatively reliable. We now argue that this dynamics generically gives a sizeable spin asymmetry.

The numerator of the expression \eq{asymm} for $A_N$ is an interference between two amplitudes with opposite helicity of the initial proton whereas the helicities of all other particles are the same (and summed over). The helicities are indicated by $\pm$ signs in Fig.~2. As we discussed in section 2, any helicity flip in the hard part of the process is suppressed by the small current quark mass. As generally required \cite{Farrar:1975yb}, the fast quark $k$ in Fig.~2 carries the proton helicity, which differs for the two amplitudes. The pion is formed through a recombination of the fast quark with an antiquark $\ell$ having helicity opposite to its own. In one of the amplitudes this requires a helicity flip at the production vertex of the antiquark (marked by a black dot). Since the flip occurs in a soft process the flip and non-flip amplitudes are of similar magnitude. 

The helicity difference of the soft antiquark $\ell$ in the two amplitudes of Fig.~2 is compensated by an orbital angular momentum factor $\ell^x+i\ell^y$, which in turn is correlated with $\kvec$ through the pion wave function. The energy difference $\Delta E$ between the pion and the quark-antiquark pair $k,\ell$ depends on the transverse momenta {\em relative to the pion direction} as in \eq{deltae}. Since the quark $k$ carries most of the momentum it moves essentially in the pion direction and thus contributes little to $\Delta E$. The transverse momentum of the antiquark \wrt the pion direction varies by roughly $\Delta\ell_\perp^2 \simeq 4\sqrt{1-x_F}\,\lqcd^2$ depending on the relative orientation of $\bs{\ell}_\perp$ and $\kvec$. While $\Delta\ell_\perp^2$ vanishes in the $x_F \to 1$ limit, it is sizeable even at $x_F \simeq 0.8$ ($4\sqrt{0.2}=1.8 $). In any case, since the soft antiquark production is a coherent part of the $p^\uparrow p \to \pi(x_F,\kvec)+X$ process the extent of its momentum correlation with the pion direction cannot be quantitatively evaluated, and may well be large.

The angular correlation between $\bs{\ell}_\perp$ and $\kvec$ is a crucial requirement of our (and, it would seem, any) approach that can produce a large SSA for $\kt \gg \lqcd$. The helicity flip needs to occur in a soft part of the process in order not to involve the current quark mass, and it must be coherent (correlated) with the hard part in order to affect the azimuthal dependence of the hard scattering. In the leading twist limit \eq{ltlimit} such coherence is absent and the expectation from, \eg, the Sivers effect is then $A_N \sim \lqcd/\kt$.

The expression \eq{asymm} for $A_N$ also requires that the helicity flip and non-flip amplitudes have a sizeable phase difference. The BB limit \eq{bblimit} is advantageous compared to the LT limit \eq{ltlimit} in this respect as well, since the production amplitudes are complex already at lowest order. In the example shown in Fig.~2 the intermediate state indicated by the dotted line is on-shell for a particular value of momentum fraction $y$. Due to the overall coherence, this value of $y$ depends on the soft parton momenta. The integration over $y$ gives the amplitude an imaginary part which is of the same order as the real part (see next section).

\vspace{1cm}

\noindent{\bf 4. A sample calculation}
\vspace{.5cm}

We evaluate the Feynman diagrams shown in Fig.~2 (which differ only in their helicity assignments) as a demonstration of the generic arguments presented in the previous section. Since the calculation involves soft processes the perturbative calculation is not quantitatively reliable. We use abelian gauge theory and do not consider other diagrams since we believe that our calculation is representative of the full answer.

We parametrize the momenta in Fig.~2 as follows,
\beqa
 p_{1} &=& \left(y p^+,0,\notvec\right) \nn\\
 p_{2} &=& \left((1-y) p^+,0,\notvec\right) \nn\\
 k &=& \left(zx_F p^+,\kvec^2/zx_F p^+,\kvec\right); \hspace{6cm}
 \kvec=\kt(\cos\phi,\sin\phi) \nn\\
 \ell &=& \left((1-z) x_F p^+,(\ellt^2+M^2)/(1-z)x_F p^+,\lvec\right); \hspace{2.95cm}
 \lvec=\ellt(\cos\psi,\sin\psi) \nn\\
 \ell_{1} &=& \left(w(1-x_F)p^+,0,\notvec\right) \nn\\
 \ell_{2} &=& \left((1-w)(1-x_F)p^+,(\ltt^2+M^2)/(1-w)(1-x_F)p^+,\lvect\right);\ \ \ 
  \bs{\ell}_{2\perp}=\ell_{2\perp}(\cos\tau_2,\sin\tau_2)\nn\\
\eeqa
The momenta $p_1,p_2$ of the (massless) quarks in the proton are taken to be collinear, as the intrinsic momentum of the proton wave function is much smaller than the hard scale $\kt$. To enable helicity flip the produced quark and antiquark ($\ell_2,\ell$) are assumed to have mass $M$. We consider the BB limit \eq{bblimit}, with $1-z$ of \order{1-x_F}, $w$ of \order{\halft}, $\ltt^2 \sim \ellt^2 \sim \lqcd^2$, and drop all subleading contributions in $1/s=1/E_{CM}^2$. We put $\lto=0$ to simplify the expressions below. 

The  helicity flip (F) and nonflip (NF) amplitudes take the form
\beq\label{ampexpr}
 \mM_{\mathrm{F/NF}} = g^6\frac{\mN_{\mathrm{F/NF}}}{\prod_{i=1}^5\left( q_i^2+\ieps\right)} 
\eeq
where the product of the five denominators is
\beqa \label{denom}
 \prod_{i=1}^5&&\hspace{-.5cm}\left( q_i^2\!+\!\ieps\right) = 
 \left(1-y\right)\left[1-z+(1-w)(1-x_F)\right]\left(1-z+1-x_F\right) \kt^2 \nn \\
&\times&\! \left[\frac{\ltt^2+M^2}{(1-w)(1-x_F)}+\frac{\ellt^2+M^2}{1-z}\right] 
\left[\frac{\ltt^2+M^2}{(1-w)(1-x_F)}+\frac{\ellt^2+M^2}{1-z}-\frac{(\lvect+\lvec)^2}{1-z+1-x_F}\right] \nn\\
&\times&\!  \left[\frac{\ltt^2+M^2}{(1-w)(1-x_F)}+\frac{\ellt^2+M^2}{1-z}-\frac{(\lvect+\lvec)^2}{1-z+(1-w)(1-x_F)}\right] \nn\\
&\times&\! \left[ y\left(\frac{\ltt^2+M^2}{(1-w)(1-x_F)}+\frac{\ellt^2+M^2}{1-z}\right)-(1-y)\kt^2 + \ieps \right]
\eeqa
The last factor corresponds to the denominator cut by the dotted line in Fig.~2 which vanishes in the range $0<y<1$. The numerators in \eq{ampexpr} are
\beqa \label{l1ampl}
 \mN_{\mathrm{F}} &=& 
 8Ms\,\frac{\sqrt{wy^3(1-y)}}{\sqrt{(1-w)(1-z)}}\,\kex{}\left(\lext{-}+\qex{-}\right)^2\nn\\ \nn\\
 \mN_{\mathrm{NF}} &=& 
 8s\, \sqrt{\frac{wy(1-y)}{(1-w)(1-z)}}\,\kex{}\Big[\left(\lext{-}+\qex{-}\right)\nn\\
&& \times\left(\lext{-}\qex{}- M^2\right) - (\ell^- + \ell_2^-)p^+(1-z)\lext{-} \Big]
\eeqa

The amplitudes should be weighted by the proton distribution amplitude and integrated over the quark momentum fraction $y$. The result is not sensitive to the shape of the distribution amplitude -- taking it to be constant the 
$y$-integrals are of the form
\beqa
 I_{\mathrm{F}}&\equiv&\int_0^1\, dy \sqrt{\frac{y^3}{1-y}}\,  \inv{yA-(1-y)B+\ieps} = \frac{\pi}{2\sqrt{A}}\frac{\sqrt{A}+2i\sqrt{B}}{(\sqrt{A}+i\sqrt{B})^2} \nn\\
 I_{\mathrm{NF}}&\equiv&\int_0^1\, dy \sqrt{\frac{y}{1-y}}\, \inv{yA-(1-y)B+\ieps} = \frac{\pi}{\sqrt{A}(\sqrt{A}+i\sqrt{B})} 
\eeqa
where
\beqa
 A&=&\frac{\ltt^2+M^2}{(1-w)(1-x)}+\frac{\ellt^2+M^2}{1-z}\nn\\
 B&=&\kt^2
\eeqa
The dynamical phase difference between the helicity flip and nonflip amplitudes which is required for $A_N$ in \eq{asymm} arises from the ratio
\beq
 \frac{I_{\mathrm{F}}}{I_{\mathrm{NF}}} = \frac{A+2B+i\sqrt{AB}}{2(A+B)} = \left|  \frac{I_{\mathrm{F}}}{I_{\mathrm{NF}}}\right| \exp(i\theta)
\eeq

where 
\beq \label{dynp}
 \tan \theta = \frac{\sqrt{AB}}{{A+2B}};\quad \sin \theta = \frac{\sqrt{AB}}{\sqrt{A^2+5AB+4B^2}}
\eeq
Outside of the scaling region \eq{bblimit} either $A/B \to 0$ or $B/A \to 0$, in which case $\theta \to 0$ and $A_N$ vanishes. This illustrates the importance of coherence between the soft and hard subprocesses.

As we discussed in section 3, $A_N \neq 0$ also requires a correlation between the azimuthal angles ($\phi$ and $\psi$) of the quark and antiquark that form the pion. While such a correlation is expected already kinematically, the coherent soft production process is likely also to depend on $\phi-\psi$. An estimate of the size of $A_N$ would thus require further, model-dependent assumptions.

\vspace{1cm}

\noindent{\bf 5. Discussion}
\vspace{.5cm}

We addressed the single spin asymmetries observed in $p^\uparrow p \to \pi(x_F,\kvec)+X$ by E704 \cite{Adams:1991rw} at high $x_F$ in the framework of the kinematical limit \eq{bblimit}, where $x_F \to 1$ as the hard scale $\kt \to \infty$. In this BB limit \cite{Berger:1979du} coherence is maintained over the entire proton distribution amplitude. Hence it differs essentially from the leading twist limit \eq{ltlimit} where $x_F$ is held fixed and single projectile partons contribute incoherently. Several aspects of the data motivated us to consider the BB limit:

\begin{itemize}
\item[--] The discrepancy between the leading twist cross section and the E704 data \cite{Bourrely:2003bw};

\item[--] The observation of coherence effects in 
the Drell-Yan process at similar $x_F$ and for a considerably harder scale \cite{Anderson:1979xx};

\item[--] The increase of the spin asymmetry with $x_F$ and $\kt$. The Sivers distribution is expected to be suppressed by a factor $1-x_F$ compared to the unpolarized quark distribution  \cite{Brodsky:2006hj}, and to give $A_N \sim \lqcd/\kt$;

\item[--] The measured $A_N \sim 0.4$, which is an order of magnitude larger than the spin asymmetry observed in DIS \cite{Airapetian:2004tw}.

\end{itemize}

The leading order amplitudes contributing to $pp \to \pi +X$ in the BB limit, such as the diagram of Fig.~2, have large imaginary parts and unsuppressed helicity flip contributions from the soft subprocesses. Because of the coherence between the soft and hard parts of the amplitude these features are directly relevant to the spin asymmetry of pions produced at high $\kt$. While the necessary ingredients for a large $A_N$ thus are present, perturbative calculations cannot give a reliable value for $A_N$ due to the importance of the soft interactions.

We focussed on the E704 data as most suitable for the limit \eq{bblimit}. Our considerations may also be relevant for the STAR data, which shows \cite{Adams:2003fx,Surrow} that $A_N$ remains sizeable up to $k_\perp \simeq 3$ GeV.
The large single spin asymmetry in $pp \to \Lambda^\uparrow(x_F,\kvec) X$ \cite{Bunce:1976yb,Lundberg:1989hw}, which does not decrease up to $\kt \simeq 3.5\ \gev$, also strongly suggests coherence effects between hard and soft subprocesses. The asymmetries in $\Lambda$ and $\Xi^0$ production have been observed to be similar  \cite{Heller:1983ia}. Hence $A_N$ appears not to be sensitive to whether the projectile and produced particle share one or two valence quarks. In $pp \to \bar\Lambda^\uparrow(x_F,\kvec) X$, on the other hand, there are no shared valence quarks and the asymmetry vanishes \cite{Lundberg:1989hw}. Conversely, in $pp \to p^\uparrow(x_F,\kvec) X$ no valence quark needs to be stopped and the proton polarization is again \cite{Polvado:1979zf} much smaller than in $\Lambda$ and $\Xi^0$ production. These features are qualitatively consistent with the dynamics discussed in this paper, which is based on a common valence quark transferring momentum and helicity  from projectile to final hadron, and helicity flip effects occurring in soft (anti)quark production.

If the approach discussed here is correct, data on single spin asymmetries may yield important information on the soft coherent dynamics that cannot be properly described using perturbation theory. An improved understanding of particle production close to the exclusive limit may via the Drell-Yan-West relation \cite{Drell:1969km} be relevant also for hard exclusive processes. In Fig.~2 the pion is initially produced in a very asymmetric state, with the quark carrying much more momentum than the antiquark. The relevance of such endpoint configurations is an important issue for the Brodsky-Lepage description \cite{Lepage:1980fj} of hard exclusive scattering. 

\acknowledgments
We wish to thank Stan Brodsky and Markus Diehl for helpful discussions. The work of PH and MJ is supported in part by the Academy of Finland through grant 102046. MJ acknowledges a PhD study grant from GRASPANP, the Finnish Graduate School in Particle and Nuclear Physics.

\end{document}